\documentclass[aps,amsmath,showpacs,amsfonts,10pt]{revtex4}
\usepackage{epsfig,graphicx}
\usepackage[english]{babel}
\usepackage{amsfonts}
\usepackage{amsmath}
\usepackage{latexsym}
\usepackage{graphics,bm}
\usepackage{natbib}
\usepackage{dcolumn}
\usepackage{bm}
\usepackage{rotating}

\begin{document}

\title{Amplitude and phase mode in a Bose Einstein condensate}

\author{Aranya B Bhattacherjee }

\address{Department of Physics, ARSD College, University of Delhi (South Campus), New Delhi-110021, India}

\begin{abstract}
We show that starting from the Heisenberg equations of motion for Bose annihilation and creation operators and using an appropriate transformation, we can split the Bogoliubov mode into a free particle mode and the amplitude mode. We show this for both the free Bose gas as well as the Bose gas in an optical lattice.
\end{abstract}


\maketitle

\section{Introduction}
The dynamical behaviour of Bose-Einstein condensate (BEC) such as collective modes is one of the important source of information about the physical characteristics of the condensate. More ever, the spectrum of elementary excitations of the condensate is utilized to derive the thermodynamic properties. The properties of elementary excitations may be investigated by considering small deviations of the state of the BEC from equilibrium and finding periodic solutions of the time-dependent Gross-Pitaeviskii equation. The resulting excitation spectrum is called the Bogoliubov spectrum which for a uniform Bose gas is $\epsilon_{q}=\sqrt{\epsilon_{q}^{0}(\epsilon_{q}^{0}+2 n_{0}U)}$ \cite{pethick}, where $n_{0}$ is the uniform density of the cloud, $U$ is the two body interaction strength and $\epsilon_{q}^{0}=\frac{\hbar^2 q^2}{2m}$ is the free particle energy with $m$ as the mass of the atom and $q$ is the wave number of the excitations. For small $q$, $\epsilon_{q}$ is a linear function of $q$ i.e $\epsilon_{q}=\sqrt{\frac{n_{0}U}{m}}\hbar q$, which is phonon like. On the other hand at short wavelength $\epsilon_{q}\simeq \epsilon_{q}^{0}+n_{0}U$. In any case $\epsilon_{q}\rightarrow 0$ as $q \rightarrow 0$, a typical sound-like behaviour. In addition, we note that the Bogliubov mode comprises of two modes, namely $\epsilon_{q}^{0}/ \hbar$ and $(\epsilon_{q}^{0}+2 n_{0}U)/ \hbar$. The first mode is simply the free particle mode while the second mode is like the much discussed amplitude mode \citep{huber1, huber2, cazalilla, axel, extra}. The amplitude mode has a property that it gives rise to a gap at $q=0$. The amplitude mode is basically the oscillations of the amplitude of the order parameter of the BEC and the phase mode is the spatial and temporal oscillations of the phase of the order parameter. Recently, the amplitude or the gapped mode has been discussed in the context of the Bose-Hubbard model \citep{huber1, huber2}. It was shown that close to the superfluid-insulator transition $U \sim t \tilde{n}$ ($t$ is the hopping parameter and $\tilde{n}$ is the mean filling per site), the combined action of the lattice and the interaction leads to a Lorentz-invariant critical theory which admits the existence of an amplitude mode. On the other hand, in the Gross-Pitavaeskii theory the density mode is bound to the phase degree of freedom, resulting in the unique and well known Bogoliubov mode. The amplitude mode was detected recently using the technique of Bragg spectroscopy \citep{ulf}. In the following, we show that a simple transformation can separate the amplitude and phase degrees of freedom. In particular we analyze the uniform Bose gas and a Bose gas confined in an optical lattice.

\section{The uniform Bose gas}

Let us consider a uniform gas of interacting bosons contained in a box of volume $V$. Within the Bogoliubov approach (equivalent to including terms which are no more than quadratic in $a_{q}$ (annihilation operator for a Bose particle with momentum $q$) and $a_{q}^{\dagger}$ (creation operator for a Bose particle with momentum $q$) ) the Hamiltonian is written as \citep{pethick}

\begin{equation}
H=\frac{N^2U}{2V}+\sum_{q(q\neq 0)} [(\epsilon_{q}^{o}+n_{0}U)(a_{q}^{\dagger} a_{q}+a_{-q}^{\dagger}a_{-q})+n_{0}U(a_{q}^{\dagger}a_{-q}^{\dagger}+a_{q}a_{-q})],
\end{equation}

where $N$ is the expectation value of $\hat{H}=\sum_{q} a_{q}^{\dagger}a_{q}$. In Eqn.(1), we have also taken into account that the total number of particles is fixed. The chemical potential is $\mu=n_{0}U$. Note that in Eqn.(1), the summation is to be taken over only one half of momentum space, since the terms corresponding to $q$ and $-q$ must be counted only once. We now write down the Heisenberg equation of motion for $a_{q}$ and $a_{-q}$ from Eqn.(1). This yields

\begin{equation}
\frac{d a_{q}}{dt}=-i \epsilon_{o} a_{q}-i \epsilon_{1} a_{-q}^{\dagger},
\end{equation}

\begin{equation}
\frac{d a_{-q}}{dt}=-i \epsilon_{o} a_{-q}-i \epsilon_{1} a_{q}^{\dagger},
\end{equation}

where $\epsilon_{o}=(\epsilon_{q}^{o}+n_{0}U )/ \hbar$ and $\epsilon_{1}=n_{0}U/ \hbar$. We now make a transformation $a_{q}\rightarrow \tilde{a}_{q} e^{i \omega t}$ and $a_{-q}\rightarrow \tilde{a}_{-q} e^{-i \omega t}$. This yields from Eqns.(2) and (3), the usual Bogoliubov mode

\begin{equation}
\omega_{Bog}=\sqrt{\tilde{\epsilon_{q}}^{o}(\tilde{\epsilon_{q}}^{o}+2 n_{0}\tilde{U})},
\end{equation}

where $\tilde{\epsilon_{q}}^{o}=\frac{\hbar q^{2}}{2 m}$ and $\tilde U=\frac{U}{\hbar}$. On the other hand a transformation of the type $a_{q}\rightarrow \tilde{a}_{q} e^{-i \omega t}$ and $a_{-q}\rightarrow \tilde{a}_{-q} e^{-i \omega t}$ yields two modes

\begin{equation}
\omega_{+}=\tilde{\epsilon_{q}}^{o}+2 n_{0} \tilde{U},
\end{equation}

\begin{equation}
\omega_{-}=\tilde{\epsilon_{q}}^{o}.
\end{equation}

The $\omega_{-}$ branch is the free particle mode while $\omega_{+}$ branch is the amplitude or the gapped branch with the property that $\omega_{+} \rightarrow 2 \tilde \mu$ as $q \rightarrow 0$, $\tilde \mu=n_{0} \tilde U$. For a Bose gas in an optical lattice it was shown that the gap is exactly $2 \tilde \mu$ \citep{axel}. The above result is independent of the fact whether $n_{0} \tilde U$ $<<$ $1$ or $ n_{o}\tilde U$ $>>$ $1$.

\begin{figure}[h]
\hspace{-0.0cm}
\begin{tabular}{cc}
\includegraphics [scale=0.65]{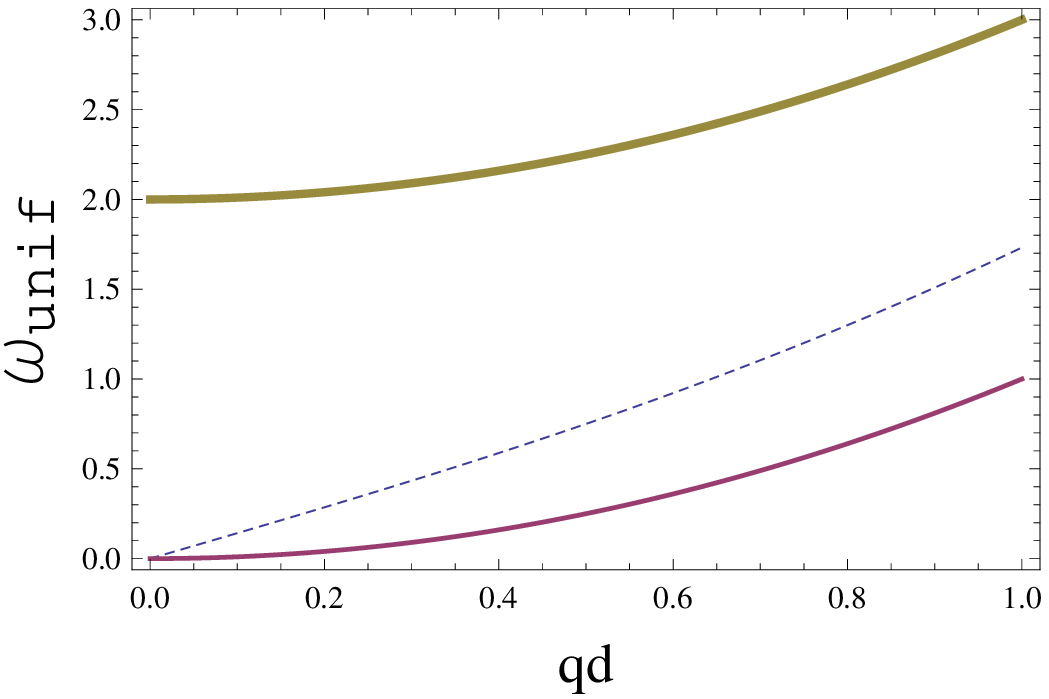}& \includegraphics [scale=0.65] {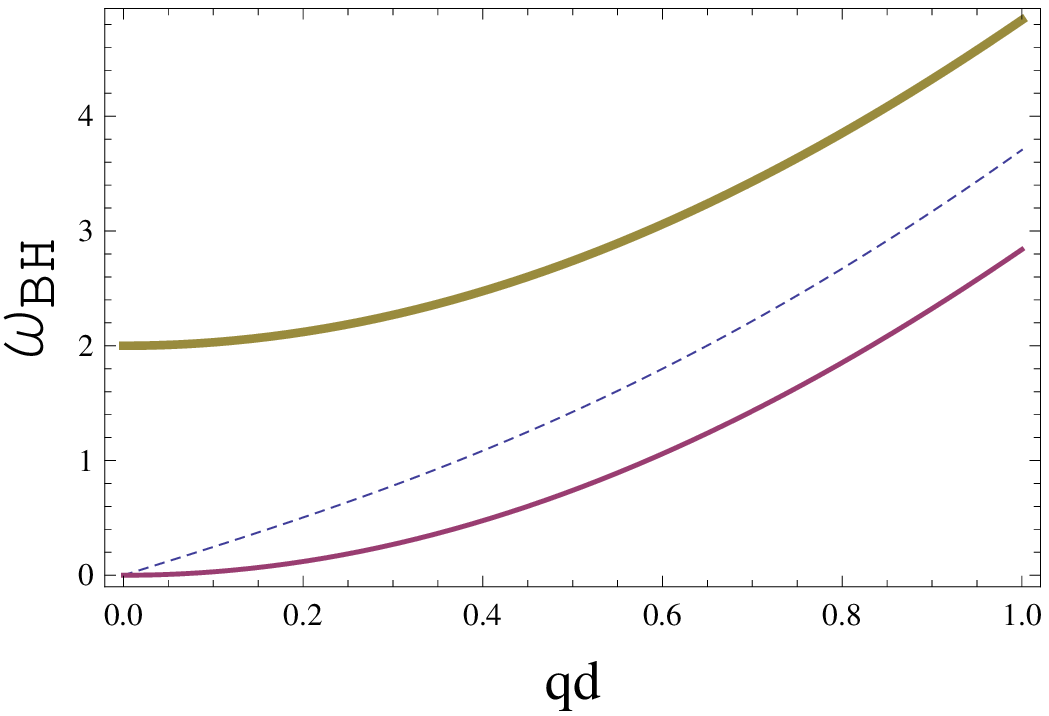}\\
 \end{tabular}
\caption{Figure depicting the amplitude mode (solid think line), free particle mode (thin solid line) and the Bogoliubov mode (dashed lined) as a function of $qd$ for the uniform Bose gas (left plot) and the Bose gas in an optical lattice (right plot). The mode frequencies are dimensionless with respect to $n_{0}\tilde{U}$ for the uniform gas and with respect to $\tilde{n} \tilde{U}$ for the Bose-Hubbard model. }
\label{f1}
\end{figure}

\section{Bose gas in an optical lattice}

We now consider a Bose gas confined in an optical lattice. We ignore the parabolic confining potential and write down the corresponding Bose-Hubbard Hamiltonian in the spirit of the Bogoliubov approximation as in the previous section

\begin{eqnarray}
H_{BH}&=&\frac{N^{2}U}{2I}+\sum_{j,q(q \neq 0)} [\epsilon_{q,j}+2 \tilde n U](a_{j,q}^{\dagger}a_{j,q}+a_{j,-q}^{\dagger} a_{j,-q}) \nonumber \\
&-& t\sum_{j,q (q \neq 0)} [(a_{j,q}^{\dagger} a_{j+1,q}+a_{j,q}^{\dagger} a_{j-1,q})+(a_{j,-q}^{\dagger} a_{j+1,-q}+a_{j,-q}^{\dagger} a_{j-1,-q})] \nonumber \\
&+& \tilde n U \sum_{j,q (q \neq 0)} (a_{j,q}^{\dagger} a_{j,-q}^{\dagger}+a_{j,q} a_{j,-q}).
\end{eqnarray}

Here $\epsilon_{j,q}$ are the onsite energies , $t$ is the hopping parameter and $I$ is the total number of sites. The Heisenberg equation of motion for $a_{j,q}$ and $a_{j,-q}$ is

\begin{equation}
\frac{d a_{j,q}}{dt}=-i \tilde {\epsilon}_{o} a_{j,q}-i \tilde{\epsilon_{1}}a_{j,-q}+i t (a_{j+1,q}+a_{j-1,q}),
\end{equation}

\begin{equation}
\frac{d a_{j,-q}}{dt}=-i \tilde {\epsilon}_{o} a_{j,-q}-i \tilde{\epsilon_{1}}a_{j,q}+i t (a_{j+1,-q}+a_{j-1,-q}),
\end{equation},

where $\tilde{\epsilon}_{o}= \epsilon_{q}+2 \tilde {n} U$ and $\epsilon_{1}= \tilde{n} U$. We have assumed $\epsilon_{j,q}=\epsilon_{q}$. Again a transformation of the type $a_{j,q}\rightarrow \tilde{a}_{j,q} e^{iqjd} e^{i \omega t} e^{-i \bar{\mu} t}$ and $a_{j,-q}\rightarrow \tilde{a}_{j,-q} e^{-iqjd} e^{-i \omega t} e^{-i \bar{\mu} t}$ ($\bar{\mu}=\tilde{n} \tilde{U}-2 \tilde{t}$ and $d$ is the lattice spacing) yields the Bogoliubov mode for the Bose-Hubbard Hamiltonian

\begin{equation}
\omega_{Bog}^{BH}=\sqrt{(\tilde{\epsilon}_{q}+4 \tilde{t} \sin^{2}{qd/2})[\tilde{\epsilon}_{q}+2 \tilde{n} \tilde{U}+4 \tilde{t} \sin^{2}{qd/2}]}.
\end{equation}

Here $\omega_{Bog}^{BH} \rightarrow 0$ as $q \rightarrow 0$. Similarly the transformations $a_{j,q}\rightarrow \tilde{a}_{j,q} e^{iqjd} e^{-i \omega t} e^{-i \bar{\mu} t}$ and $a_{j,-q}\rightarrow \tilde{a}_{j,-q} e^{-iqjd} e^{-i \omega t} e^{-i \bar{\mu} t}$ splits the Bogoliubov mode into two modes

\begin{equation}
\omega_{+}^{BH}= \tilde{\epsilon}_{p}+2 \tilde{n} \tilde{U}+4 \tilde{t} \sin^{2}{qd/2},
\end{equation}

\begin{equation}
\omega_{-}^{BH}= \tilde{\epsilon}_{p}+4 \tilde{t} \sin^{2}{qd/2},
\end{equation}

The amplitude mode $\omega_{+}^{BH}$ $\rightarrow$ $2 \tilde{n} \tilde{U}$  as $q \rightarrow 0$.  Note that the above theory makes use of the "breaking of symmetry" in frequency space to separate the amplitude mode from the phase mode. Take for example the uniform Bose gas. The terms $(a_{q}^{\dagger} a_{q}+a_{-q}^{\dagger}a_{-q})$ and $(a_{q}^{\dagger}a_{-q}^{\dagger}+a_{q}a_{-q})$ evolve under the transformation $a_{q}\rightarrow \tilde{a}_{q} e^{i \omega t}$ and $a_{-q}\rightarrow \tilde{a}_{-q} e^{-i \omega t}$ as $(\tilde{a}_{q}^{\dagger} \tilde{a}_{q}+ \tilde{a}_{-q}^{\dagger}\tilde{a}_{-q})$ and $(\tilde{a}_{q}^{\dagger}\tilde{a}_{-q}^{\dagger}+\tilde{a}_{q}\tilde{a}_{-q})$. On the other hand, the transformations $a_{q}\rightarrow \tilde{a}_{q} e^{-i \omega t}$ and $a_{-q}\rightarrow \tilde{a}_{-q} e^{-i \omega t}$ yield $(\tilde{a}_{q}^{\dagger} \tilde{a}_{q}+ \tilde{a}_{-q}^{\dagger}\tilde{a}_{-q})$ and $(\tilde{a}_{q}^{\dagger}\tilde{a}_{-q}^{\dagger} e^{2i \omega t}+\tilde{a}_{q}\tilde{a}_{-q}e^{-2i \omega t})$, indicating a breaking of symmetry due to a two-photon process. Such asymmetry can be created by exposing the BEC to two counter-propagating Bragg pulses such that the atoms undergo two-photon process and the amplitude mode can be detected only at higher Bragg frequency \citep{ulf}.

\section{Acknowledgements}

The author thanks Axel Pelster for some stimulating discussions on amplitude mode.

\end{document}